\documentclass[onecolumn,showpacs,preprintnumbers,showkeys]{revtex4}
\usepackage{amsthm, amscd, amsfonts, amssymb, graphicx}
\usepackage{dcolumn}
\usepackage{bm}
\usepackage[english]{babel}
\usepackage{footnpag}
\usepackage{nccfoots}
\usepackage[symbol*]{footmisc}

\begin{document}

\title{The role of phase fluctuations in model of superconductivity with the external pair potential}
\author{Konstantin V. Grigorishin}
\email{konst.phys@gmail.com} \affiliation{Boholyubov Institute for
Theoretical Physics of the National Academy of Sciences of
Ukraine, 14-b Metrolohichna str. Kiev-03680, Ukraine.}
\date{\today}

\begin{abstract}
We investigate the stability of the solutions of the BCS model
with the external pair potential formulated in a work \emph{K.V.
Grigorishin} arXiv:1605.07080. It has been shown that the cause of
superconductivity in this model is the electron-electron
interaction only and the potential cannot impose superconductivity
since the superconducting ordering will be destroyed by phase
fluctuations in a form of quantum turbulence (vortexes, vortex
rings, vortex tangles etc.). The external pair potential
renormalizes the order parameter in initial superconducting matter
and the bosonic scenario of transition to normal state in a clear
$3D$ superconductor occurs unlike BCS theory where the fermionic
scenario takes place.
\end{abstract}

\keywords{external pair potential, order parameter, topological
defect, quantum turbulence}

\pacs{74.20.Fg, 74.20.Mn}

\maketitle

\section{Introduction}\label{intr}

According to BCS theory the transition to superconducting state is
accompanied by reconstruction of the electron spectrum with
appearance of the energy gap $2|\Delta|$ on Fermy level. The
superconducting state is characterized by the complex order
parameter
\begin{equation}\label{0.1}
    \Delta(\textbf{r})=|\Delta|e^{i\phi(\textbf{r})}
\end{equation}
whose modulus is the binding energy of a Cooper pair, and whose
phase $\phi(\textbf{r})$ characterizes the coherence of ensemble
of the Cooper pairs. When there is no current in the
superconductor then $\phi(\textbf{r})=const$. Two scenarios of
superconductor-normal metal transition exist. In BCS theory the
module $|\Delta|$ becomes zero in the point of phase transition
$|\Delta(T_{c})|=0$, and the phase loses its sense automatically.
This is fermionic scenario of the transition. However
superconducting state can be destroyed with another way:
fluctuations of order parameter turns to zero the correlator:
\begin{equation}\label{0.2}
    \left\langle\Delta^{+}(0)\Delta(\textbf{r})\right\rangle\rightarrow
    0 \quad \texttt{for}\quad r\rightarrow\infty,
\end{equation}
at nonzero module of order parameter $|\Delta|\neq 0$. This is
bosonic scenario of the transition. The finite module at
transition means presence of finite concentration of bound
electron pairs, i.e. concentration of bosons in the transition
does not become zero. Such regime can be released in system with
low electron density due small "rigidity" with respect to the
phase's changes \cite{emery1,emery2}. The bosonic scenario can
occur in granular superconductors where Cooper pairs localized in
granules so that superconductivity of macroscopic sample disappear
but the granules remain superconducting \cite{gant1,gant2,lar}.
The bosonic scenario takes place in $2D$ superconductivity
\cite{gant2}: the two-dimensional superconducting systems are
distinguished by the possible existence of a gas of fluctuations
in the form of spontaneously generated magnetic vortices below the
bulk transition temperature $T_{c0}$. The vortices are created in
pairs with opposite directions of the on-axis field
(vortex-antivortex pairs). A transit around a vortex changes the
phase of the wave function by $2\pi$, so that free movement of the
vortices leads to fluctuations in the phase. If the amplitude of
the fluctuations in the phase is sufficiently high, the coherence
of the state is lost $\left\langle
e^{i\phi(\textbf{r})}\right\rangle=0$. However the modulus of the
order parameter remains nonzero over a larger part of the volume
$\left\langle|\Delta(\textbf{r})|\right\rangle\neq0$ (it goes to
zero only inside vortices, near the vortex axis). Within some
range of temperatures $T_{\texttt{BKT}}<T<T_{c0}$ Cooper pairs
coexist with vortices. The pairs reduce the dissipation, but do
not suppress it completely. For temperatures $T<T_{\texttt{BKT}}$
the correlator (\ref{0.2}) tends to zero according to a power law
$1/r$, thus quasi-long-rage order and almost coherent state takes
place. It should be noted the scenario of preformed pairs
\cite{emery2,franz,cap}: superconductivity is destroyed due to
phase fluctuations at temperature $T=T_{c}$ but Cooper pairs
survive up to temperature $T^{*}$ called the pseudogap temperature
well above $T_{c}$, which manifest as the energy gap in the
electronic DOS although superconductivity has been destroyed.

In a work \cite{grig} the model of a hypothetical superconductor
has been proposed. In this model an interaction energy between
(within) structural elements of condensed matter (molecules,
nanoparticles, clusters, layers, wires etc.) depends on state of
Cooper pairs: if the pair is broken, then energy of the molecular
system is changed by quantity $\upsilon=E_{a}-E_{b}$, where
$E_{a}$ and $E_{b}$ are energies of the system after- and before
breaking of the pair accordingly. Thus to break the Cooper pair we
must make the work against the effective electron-electron
attraction and must change the energy of the structural elements:
\begin{equation}\label{1.1}
    2|\Delta|\longrightarrow 2|\Delta|+\upsilon.
\end{equation}
We will consider a case $\upsilon>0$ only: the breaking of the
pair raises of the molecular structure (creation of the pair
lowers the energy). In this case the pairs become more stable. The
Hamiltonian corresponding to the transformations (\ref{1.1}) is
\begin{eqnarray}\label{1.2}
    \widehat{H}&=&\widehat{H}_{\texttt{BCS}}+\widehat{H}_{\upsilon}=\sum_{\textbf{k},\sigma}\varepsilon(k)a_{\textbf{k},\sigma}^{+}a_{\textbf{k},\sigma}
    -\frac{\lambda}{V}\sum_{\textbf{k},\textbf{p}}a_{\textbf{p}\uparrow}^{+}a_{-\textbf{p}\downarrow}^{+}a_{-\textbf{k}\downarrow}a_{\textbf{k}\uparrow}
    -\frac{\upsilon}{2}\sum_{\textbf{k}}\left[\frac{\Delta}{|\Delta|}a_{\textbf{k}\uparrow}^{+}a_{-\textbf{k}\downarrow}^{+}
    +\frac{\Delta^{+}}{|\Delta|}a_{-\textbf{k}\downarrow}a_{\textbf{k}\uparrow}\right],
\end{eqnarray}
where $\widehat{H}_{\texttt{BCS}}$ is BCS Hamiltonian: kinetic
energy + pairing interaction ($\lambda>0$). The combinations
$a_{\textbf{k}\uparrow}^{+}a_{-\textbf{k}\downarrow}^{+}$ and
$a_{-\textbf{k}\downarrow}a_{\textbf{k}\uparrow}$ are creation and
annihilation of Cooper pairs operators. The anomalous averages
\begin{eqnarray}\label{1.3}
    \Delta^{+}=\frac{\lambda}{V}\sum_{\textbf{p}}\left\langle
    a_{\textbf{p}\uparrow}^{+}a_{-\textbf{p}\downarrow}^{+}\right\rangle,
    \quad
    \Delta=\frac{\lambda}{V}\sum_{\textbf{p}}\left\langle
    a_{-\textbf{p}\downarrow}a_{\textbf{p}\uparrow}\right\rangle,
\end{eqnarray}
are the complex order parameter $\Delta=|\Delta|e^{i\phi}$ (or
two-component order parameter: module $|\Delta|$ and phase
$\phi$). Due the multipliers $\frac{\Delta}{|\Delta|}$ and
$\frac{\Delta^{+}}{|\Delta|}$ in $\widehat{H}_{\upsilon}$ the
energy does not depend on the phase $\phi$ ($a\rightarrow
ae^{i\phi/2},a^{+}\rightarrow a^{+}e^{-i\phi/2}\Longrightarrow
\Delta\rightarrow\Delta e^{i\phi},\Delta^{+}\rightarrow
\Delta^{+}e^{-i\phi}$). Thus both $\widehat{H}_{\texttt{BCS}}$ and
$\widehat{H}_{\upsilon}$ are invariant under the $U(1)$
transformation. The term $\widehat{H}_{\upsilon}$ is similar to
"source term" in \cite{matt}, where it means the injection of
Cooper pairs into the system. On the other hand,
$\widehat{H}_{\upsilon}$ has a form of an external field acting on
a Cooper pairs only, and $\upsilon$ is energy of a Cooper pair in
this field. We call the field $\upsilon$ as the \textit{external
pair potential} (EPP) since the potential is imposed on the
electron subsystem by the structural elements of matter. It has
been obtained two type of solutions in this model (at large
temperatures $T\rightarrow\infty$):
\begin{eqnarray}
|\Delta|&=&\frac{\upsilon}{2}\quad\texttt{if}\quad g=0\label{1.4}\\
|\Delta|&=&\frac{g\hbar\omega\upsilon}{4k_{B}T},\label{1.5}
\end{eqnarray}
where $\omega$ is the phonon frequency. We can see that in the
first case the energy gap is imposed on the system by EPP and it
does not depend on temperature. In the second case the energy gap
asymptotically tends to zero when temperature rises and $\Delta=0$
if electron-phonon interaction is absent $g=0$.

In this work we consider a question: what is cause of
superconductivity in the model (\ref{1.2}) - the electron-electron
interaction $\lambda$ or the potential $\upsilon$. In other words:
can we impose the superconducting state on the noninteracting
system by EPP. This question is closely related with scenario of
the superconductor-normal conductor transition: in BCS theory
(when $\upsilon=0$) we have the fermionic scenario, however
presence of EPP can change situation towards the bosonic scenario
due important role of phase fluctuations in such system. In this
work we explore the phase fluctuations and topological defects
(vortexes, vortex rings, vortex tangles and other form of quantum
turbulence) of the order parameter $\Delta(\textbf{r})$ and their
essential difference from the same in Ginzburg-Landau (GL) theory.

\section{Two types of solutions}\label{types}

In order to find equilibrium value of the order parameter $\Delta$
we should obtain system's energy. Using the BCS wave function
$|BCS\rangle=\prod_{\textbf{k}}\left(u(k)+\textsl{v}(k)a_{\textbf{p}\uparrow}^{+}a_{-\textbf{p}\downarrow}^{+}\right)|0\rangle$,
where $u^{2}(k)+\textsl{v}^{2}(k)=1$, we obtain the energy in a
form:
\begin{equation}\label{2.1}
W_{s}=\langle\widehat{H}\rangle=\sum_{\textbf{k}}2\varepsilon(k)\textsl{v}^{2}(k)-\frac{\lambda}{V}\sum_{\textbf{k},\textbf{p}}
u(k)\textsl{v}(k)u(p)\textsl{v}(p)-\upsilon\frac{\Delta}{|\Delta|}\sum_{\textbf{k}}u(k)\textsl{v}(k),
\end{equation}
where $\Delta$ can be supposed real in the absence of magnetic
field and current. The first term is internal kinetic energy, the
second term is energy of the electron-electron interaction, the
third term is energy of Cooper pairs' condensate in the external
pair field. The functions $\textsl{v}(k)$ and $u(k)$ have to
minimize the energy, that is $\frac{dW}{d\textsl{v}^{2}(k)}=0$:
\begin{equation}\label{2.2}
2\varepsilon(k)-\frac{1-2\textsl{v}^{2}(k)}{u(k)\textsl{v}(k)}\left[\frac{\lambda}{V}\sum_{\textbf{p}}
u(p)\textsl{v}(p)+\frac{\upsilon}{2}\frac{\Delta}{|\Delta|}\right]=0.
\end{equation}

There are two different solutions of this equation, which
correspond two different physical situations:

\subsubsection{Regime of withholding of the energy gap}\label{hold}

We can suppose
\begin{equation}\label{2.3}
\Delta=\frac{\lambda}{V}\sum_{\textbf{p}}u(p)\textsl{v}(p)+\frac{\upsilon}{2}\frac{\Delta}{|\Delta|}.
\end{equation}
Then we obtain the $u,\textsl{v}$ functions in a form
\begin{equation}\label{2.4}
\textsl{v}^{2}(k)=\frac{1}{2}\left(1-\frac{\varepsilon(k)}{E}\right)\quad
u^{2}(k)=\frac{1}{2}\left(1+\frac{\varepsilon(k)}{E}\right)
\end{equation}
as in BCS theory with
\begin{equation}\label{2.5}
E^{2}=\varepsilon^{2}+|\Delta|^{2},\quad
\textsl{v}(k)u(k)=\frac{\Delta}{2E}.
\end{equation}
From Eq.(\ref{2.3}) we can see that if the coupling constant is
$\lambda=0$ then the energy gap is nonzero $|\Delta|=\upsilon/2$
(if $\upsilon>0$). Thus the operator $\widehat{H}_{\upsilon}$
describes the external source of Cooper pairs injecting the pairs
into the system. If to suppose $\upsilon=0$ then we have ordinary
BCS theory.

Superconducting state is energetically favorable if the free
energy (\ref{2.1}) is less than the free energy of normal state
i.e $W_{s}-W_{n}<0$. Using method of a work \cite{grig} we can
obtain the free energy $F_{s}-F_{n}$, where if we suppose the
absence of EPP $\upsilon=0$ and $T\rightarrow T_{c}$ then we have
GL expansion:
\begin{equation}\label{2.8}
    F_{s}-F_{n}=-a|\Delta|^{2}+\frac{b}{2}|\Delta|^{4}+cq^{2}|\Delta|^{2},
\end{equation}
where $a=\alpha(T_{c}-T),\alpha>0$, $b$ and $c$ are expansion
coefficients. The equilibrium value of the energy gap is
\begin{equation}\label{2.9}
    |\Delta|^{2}=\frac{a-cq^{2}}{b}.
\end{equation}
If we suppose the absence of electron-electron interaction
$\lambda=0$ then at $T\rightarrow\infty$ it is easy to obtain
\begin{equation}\label{2.10}
    F_{s}-F_{n}=-\upsilon\frac{w}{2k_{B}T}|\Delta|+\frac{w}{2k_{B}T}|\Delta|^{2}+\frac{wk_{F}^{2}}{144(k_{B}T)^{2}m^{2}}q^{2}|\Delta|^{2}
    \equiv-a|\Delta|+\frac{b}{2}|\Delta|^{2}+\frac{c}{2}q^{2}|\Delta|^{2},
\end{equation}
where $w$ is a value of energy dimension, $a$, $b$ and $c$ are
corresponding expansion coefficients. The equilibrium value of the
energy gap is
\begin{equation}\label{2.11}
    |\Delta|=\frac{a}{b+cq^{2}}=\frac{\upsilon/2}{1+\frac{c}{b}q^{2}}.
\end{equation}
We can see fundamentally differences of the energy gap (\ref{2.9})
from the energy gap (\ref{2.11}):
\begin{enumerate}
    \item the gap (\ref{2.9}) depends on temperature and $\Delta=0$ at
    $T>T_{c}$. The gap (\ref{2.11}) does not depend on temperature: $|\Delta|=\upsilon/2=\texttt{const}$
    \item for the gap (\ref{2.9}) the critical momentum $q_{c}=a/c$ exists
    such that $\Delta=0$ at $q>q_{c}$. The gap (\ref{2.11}) tend to zero asymptotically $\Delta(q\rightarrow\infty)\rightarrow0$,
    that is it does not become zero for any momentum.
\end{enumerate}
Thus EPP imposes superconductivity to the system and holds it
hardly. In the same time the coherence length is determined as
\begin{equation}\label{2.11a}
   \xi^{2}/\hbar^{2}=
   \begin{array}{ccc}
     \frac{c}{a}\propto\frac{1}{|T-T_{c}|} &  & \texttt{GL theory} \\
     & &\\
     \frac{c}{b}\propto\frac{1}{T} &  & \texttt{the withholding of the energy gap} \\
   \end{array}
\end{equation}
The coherence length has physical sense of the spatial scale at
which the magnitude of the energy gap $|\Delta|$ varies (below
critical temperature) or it is a correlation radius of
fluctuations (above the critical temperature). In the GL theory
$\xi\propto\frac{\hbar}{q_{c}}$, but in the regime of the
withholding the coherence length is finite quantity, although
$q_{c}=\infty$ formally. This fact is a consequence of the fact
that in the GL theory on a spatial scale less than $\xi$ the
energy gap is filled with quasiparticles with energy scale
$|\Delta|^{2}/E_{F}\ll |\Delta|$ \cite{genn}, for example, each
vortex line is equivalent to a normal region of radius $\xi$. In
the regime with free energy (\ref{2.10}) the EPP withhold the
energy gap on all spatial scales.

\subsubsection{Regime of renormalization}\label{ren}
On the other hand, in order to obtain the functions (\ref{2.4}) we
can suppose
\begin{equation}\label{2.12}
\Delta=\frac{\lambda}{V}\sum_{\textbf{p}}u(p)\textsl{v}(p),
\end{equation}
however then we will have a dispersion law of quasiparticles and a
condensation amplitude in forms
\begin{equation}\label{2.13}
E^{2}=\varepsilon^{2}+|\Delta|^{2}\left(1+\frac{\upsilon}{2|\Delta|}\right)^{2},\quad
\textsl{v}(k)u(k)=\frac{\Delta\left(1+\frac{\upsilon}{2|\Delta|}\right)}{2E}.
\end{equation}
Even if the ordering $\Delta\sim\left\langle aa\right\rangle,
\Delta^{+}\sim\left\langle a^{+}a^{+}\right\rangle$ is absent then
the energy gap in the quasiparticles' specter is present. Thus the
state with $\Delta=0$ can be interpreted as state with a
pseudogap. The integration domain in the self-consistency equation
(\ref{2.12}) must be cut off at some characteristic phonon energy
$\hbar\omega$ ($\lambda>0$ if $|\varepsilon(k)|<\hbar\omega$,
outside $\lambda=0$), then we have
\begin{equation}\label{2.14}
\Delta=g\int_{-\hbar\omega}^{\hbar\omega}d\varepsilon
\frac{\Delta\left(1+\frac{\upsilon}{2|\Delta|}\right)}{2\sqrt{\varepsilon^{2}+|\Delta|^{2}\left(1+\frac{\upsilon}{2|\Delta|}\right)^{2}}}.
\end{equation}
We can see from Eq.(\ref{2.14}), that if the coupling constant is
$g\equiv\lambda\nu_{F}=0$ then $\Delta=0$. This means, that only
electron-electron coupling is the cause of superconductivity but
not the potential $\upsilon$. In this case the operator
$\widehat{H}_{\upsilon}$ is an external field acting on the Cooper
pairs only, and $\upsilon$ is energy of the pair in this field. As
stated above, the potential $\upsilon$ is called as
\textit{external pair potential} since it is imposed on the
electron subsystem by the structural elements of a substance
(molecular, clusters etc. if their energy depends on state of
Cooper pairs), unlike the pair potential $\Delta$, which is result
of electron-electron interaction and determined with the
self-consistency equation. If $\upsilon=0$ we have usual
self-consistency equation $\Delta=I(\Delta)$ in BCS theory. In
presence of the external pair potential $\upsilon\neq 0$ the
self-consistency condition has a form $\Delta=I(\Delta,\upsilon)$
- Eq(\ref{2.14}). Thus the potential $\upsilon$
\emph{renormalizes} the order parameter $\Delta$.

\begin{figure}[h]
\includegraphics[width=7.5cm]{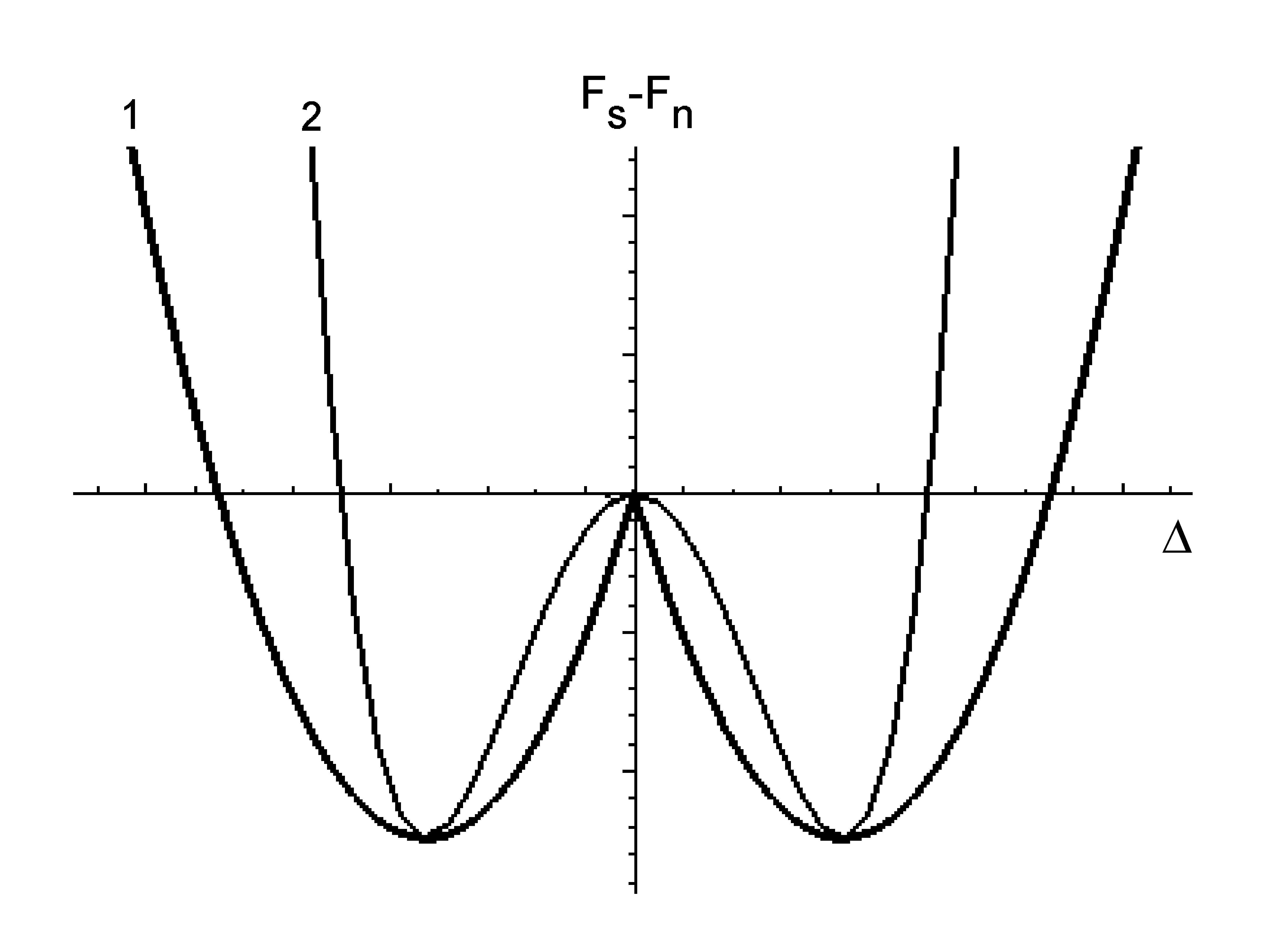}
\caption{the free energy (\ref{2.18}) at $\textbf{q}=0$ - line
(1), the Ginzburg-Landau free energy (\ref{2.8}) at $\textbf{q}=0$
- line (2).} \label{Fig1}
\end{figure}

It should be noticed that $\Delta=0$ is not solution of the
self-consistency equation (\ref{2.14}) if $\upsilon\neq 0,g\neq0$
(due the term $\frac{\Delta}{|\Delta|}\frac{\upsilon}{2}$ an
uncertainty $\frac{0}{0}$ occurs) unlike BCS theory (when
$\upsilon=0$). The absence of this solution is consequence of the
fact that the derivative $\frac{dW}{d\Delta}$ does not exist at
$\Delta=0$ because $W(\Delta\rightarrow 0)\sim |\Delta|$ and the
function $|\Delta|$ is non-differentiable in this point (the
derivative has a jump discontinuity), unlike BCS theory where
$W(\Delta\rightarrow 0)\sim |\Delta|^{2}$ - Fig.\ref{Fig1}. Thus
the regions, where Eq.(\ref{2.14}) has not solutions, should be
considered as the normal state of a superconductor. However we can
extend the definition of the normal state by the following way.
The order parameter is a complex quantity:
$\Delta=|\Delta|e^{i\phi}$. As it will be demonstrated below the
ratio $\frac{\Delta}{|\Delta|}$ can be zero due the strong
fluctuations of the phase $\phi(\textbf{r},t)$ so that average
value $\langle e^{i\phi(\textbf{r},t)}\rangle=0$. In this sense
$\Delta=0$ can be considered as a solution of the self-consistency
equation (\ref{2.14}) and it means the normal state of a
superconductor. Then the energy $W_{n}$ can be determined as
$W_{n}=W_{s}(\Delta=0)$:
\begin{eqnarray}\label{2.15}
W_{s}-W_{n}&=&V\nu_{F}\left[\int_{-\hbar\omega}^{\hbar\omega}2\varepsilon
\textsl{v}^{2}d\varepsilon-g\left(\int_{-\hbar\omega}^{\hbar\omega}\textsl{v}ud\varepsilon
\right)^{2}-\upsilon\frac{\Delta}{|\Delta|}\int_{-\hbar\omega}^{\hbar\omega}\textsl{v}ud\varepsilon\right]\nonumber\\
&-&V\nu_{F}\left[\int_{-\hbar\omega}^{\hbar\omega}2\varepsilon
\textsl{v}_{0}^{2}d\varepsilon-g\left(\int_{-\hbar\omega}^{\hbar\omega}\textsl{v}_{0}u_{0}d\varepsilon
\right)^{2}-\upsilon\int_{-\hbar\omega}^{\hbar\omega}\textsl{v}_{0}u_{0}d\varepsilon\right],
\end{eqnarray}
where
\begin{equation}\label{2.16}
\textsl{v}^{2}_{0}(\varepsilon)=\frac{1}{2}\left(1-\frac{\varepsilon}{E_{0}}\right),\quad
u^{2}_{0}(\varepsilon)=\frac{1}{2}\left(1+\frac{\varepsilon}{E_{0}}\right),\quad
u_{0}(\varepsilon)\textsl{v}_{0}(\varepsilon)=\frac{\upsilon/2}{2E_{0}},\quad
E_{0}^{2}=\varepsilon^{2}+\left(\frac{\upsilon}{2}\right)^{2}.
\end{equation}
This renormalization of the normal state is consequence of the
extrapolation in the point $\Delta=0$ where the self-consistent
equation (\ref{2.14}) does not have solutions. The expressions for
the energies $W_{s}$ and $W_{n}$ have not any sense in themselves,
only their difference $W_{s}-W_{n}$ is a physical quantity.
Interpretation for $W_{n}$ and $E_{0}$ is following. Let the
electron-electron interaction is switched off: $g=0$. Then charge
is still carried by the pairs of electrons (current carriers are
the pairs) because for their breaking it must be made the work
$\upsilon>0$. This means that the quasiparticles' spectrum has a
gap $\upsilon/2$. But this state is not superconducting because
the ordering $\left\langle aa\right\rangle,\left\langle
a^{+}a^{+}\right\rangle$ is absent. In our opinion such state can
be interpreted as state with a pseudogap.

Using method of a work \cite{grig} we can obtain the free energy
$F_{s}-F_{n}$. If the external pair potential is absent
$\upsilon=0$ and temperature is slightly less than the critical
temperature $T\lesssim T_{c}$, then we have GL expansion
(\ref{2.8}). Let consider the case $\upsilon>0$ and limit
$T\rightarrow\infty$. Then an expansion in powers of $1/T$ and
$|\Delta|$ has a form:
\begin{equation}\label{2.18}
F_{s}-F_{n}=V\nu_{F}\left[\frac{\hbar\omega}{2k_{B}T}|\Delta|^{2}-\upsilon
g\left(\frac{\hbar\omega}{2k_{B}T}\right)^{2}|\Delta|+\upsilon\frac{\hbar\omega
k_{F}^{2}}{144\left(k_{B}T\right)^{3}m^{2}}q^{2}|\Delta|\right].
\end{equation}
From condition $\frac{dF}{d|\Delta|}=0$ we obtain
\begin{equation}\label{2.19}
    |\Delta|=\frac{g\hbar\omega\upsilon}{4k_{B}T}-\frac{\upsilon k_{F}^{2}}{36(k_{B}T)^{2}m^{2}}q^{2}.
\end{equation}
Thus at large temperatures $T\gg T_{c}(\upsilon=0)$ the energy gap
tends to zero asymptotically as $1/T$ with increasing of
temperature. Thus, formally, the critical temperature is equal to
infinity. However the critical momentum
\begin{equation}\label{2.20}
    q_{c}^{2}=\frac{9g\hbar\omega m^{2}k_{B}T}{k_{F}^{2}}
\end{equation}
exists such that $\Delta=0$ at $q>q_{c}$. In \cite{grig} it has
been shown that the free energy (\ref{2.18}) is equivalent to GL
free energy (\ref{2.8}) with corresponding coefficients with $a>0$
at all temperatures, and the coherence length
\begin{equation}\label{2.21}
   \xi^{2}=\hbar^{2}\frac{c}{a}\propto\frac{1}{T}.
\end{equation}

The energy gap $\upsilon/2$ in the normal state cannot be
destroyed by the condensate's momentum $q$ as in previous regime
of withholding of the energy gap. Thus we can see that in any
given variants (regime of withholding of the energy gap or the
regime of renormalization) EPP causes the existence of the energy
gap $\upsilon/2$. The questions arise: which state can be realized
as superconducting and why the state with the gap $\upsilon/2$ in
the regime of renormalization is normal.

\section{Phase fluctuations}\label{phase}

In the mean field approximation the solution (\ref{2.9}) minimizes
the GL free energy in the superconducting phase. However the order
parameter is a complex function and the phase depends on position:
\begin{equation}\label{3.1}
    \Delta=|\Delta|e^{i\phi(\textbf{r})}.
\end{equation}
Inserting this approximation into the GL free energy we get
\begin{eqnarray}\label{3.2}
F_{s}-F_{n}&=&\int\left[-a|\Delta(\textbf{r})|^{2}+\frac{b}{2}|\Delta(\textbf{r})|^{4}
+c|\left(-i\hbar\nabla\right)\Delta(\textbf{r})|^{2}\right]d\textbf{r}\nonumber\\
&=&\texttt{const}+c|\Delta|^{2}\hbar^{2}\int|\nabla\phi(\textbf{r})|^{2}
=\texttt{const}+Vc|\Delta|^{2}\sum_{\textbf{q}}q^{2}\phi^{2}(\textbf{q}),
\end{eqnarray}
where we neglect the fluctuations of the order parameter's modulus
and the transformations
$\Delta(\textbf{r})=\sum_{\textbf{q}}\Delta_{\textbf{q}}e^{i\textbf{qr}/\hbar}$,
$\Delta_{\textbf{q}}=\frac{1}{V}\int\Delta(\textbf{r})e^{-i\textbf{qr}/\hbar}d\textbf{r}$
have been used. With the help of the equipartition theorem
$Vc|\Delta|^{2}q^{2}\left\langle\phi^{2}(\textbf{q})\right\rangle=\frac{1}{2}k_{B}T$
we obtain a Fourier component of the mean square of phase
fluctuation
\begin{equation}\label{3.3}
    \left\langle\phi^{2}(\textbf{q})\right\rangle=\frac{k_{B}T}{2c|\Delta|^{2}q^{2}V}.
\end{equation}
Let us consider the correlation function
\begin{eqnarray}\label{3.4}
    &&\left\langle\Delta^{+}(0)\Delta(\textbf{r})\right\rangle=
    |\Delta|^{2}\left\langle\exp\left(i\phi(0)-i\phi(\textbf{r})\right)\right\rangle\nonumber\\
    &&|\Delta|^{2}\left\langle\exp\left(i\sum_{\textbf{q}}\left[1-e^{i\frac{\textbf{qr}}{\hbar}}\right]\phi_{\textbf{q}}\right)\right\rangle
    \approx|\Delta|^{2}\exp\left(-\frac{1}{2}\sum_{\textbf{q}}\left|1-e^{i\frac{\textbf{qr}}{\hbar}}\right|^{2}\left\langle\phi^{2}_{\textbf{q}}\right\rangle\right)\nonumber\\
    &&=|\Delta|^{2}\exp\left(-\frac{k_{B}T}{2c|\Delta|^{2}}\frac{1}{(2\pi\hbar)^{D}}\int
    \frac{d^{D}q}{q^{2}}\left(1-\cos{\frac{\textbf{qr}}{\hbar}}\right)\right).
\end{eqnarray}
For $D=1$ we have
\begin{equation}\label{3.5}
    \int_{0}^{q_{c}}\frac{dq}{q^{2}}\left(1-\cos{\frac{\textbf{qr}}{\hbar}}\right)
    =\frac{r}{\hbar}\left(-\frac{1}{x}-\frac{\cos
    x}{x}-\texttt{Si}(x)\right)_{0}^{q_{c}r/\hbar}=+\infty\Longrightarrow
    \left\langle\Delta^{+}(0)\Delta(\textbf{r})\right\rangle=|\Delta|^{2}e^{-\infty
    r}=0
\end{equation}
This means that despite we have nonzero energy gap in each point
$|\Delta|\neq0$ the mean over system order parameter is absent
$\langle\Delta\rangle=0$. In $1D$ case superconductivity is
impossible due an \emph{infrared} divergence (at the lower limit
$0$). Physically the mechanism of destruction of phase coherence
in $1D$ systems is spontaneous creation of topological defects as
the phase slip \cite{ander,lar,tinh}. It should be noticed that
the upper limit of integration over modulus of momentum $q$ in
Eq.(\ref{3.4}) is the critical momentum $q_{c}$. This is
consequence of suppression of the energy gap such that
$|\Delta(q)|=0$ at $q>q_{c}$ - Eqs.(\ref{2.9},\ref{2.20}).

For $D=2$ it is easy to show
\begin{equation}\label{3.6}
\left\langle\Delta^{+}(0)\Delta(\textbf{r})\right\rangle=|\Delta|^{2}\left(\frac{\hbar}{rq_{c}}\right)^{\frac{k_{B}T}{4\pi\hbar^{2}
c|\Delta|^{2}}}
\end{equation}
at $r\rightarrow\infty$. Such behavior of the correlator at large
distances is known as quasi-long range order. As well known in the
normal state the correlator is
$\left\langle\Delta^{+}(0)\Delta(\textbf{r})\right\rangle\propto\frac{1}{r}e^{-r/\xi}$.
At $T=T_{c}$ we have $\xi=\infty$, thus at transition to
superconducting state the correlator becomes $\propto\frac{1}{r}$.
Hence $2D$ system is superconducting if
\begin{equation}\label{3.7}
\frac{k_{B}T}{4\pi\hbar^{2} c|\Delta|^{2}}\leq 1.
\end{equation}
Such a transition is called the BKT transition. Physically the
mechanism of destruction of phase coherence in $2D$ system is
spontaneous creation of topological defects such as vortexes
\cite{ander,lar}. The phase of order parameter
$\phi(\textbf{r},t)$ in each point of superconductor depends
strongly (in each time moment) on position of all vortexes because
the phase changes to $2\pi$ at going around each vortex. This
means, that vortexes, moving like particles of liquid, leads to
strong fluctuations of the phase $\phi(\textbf{r},t)$ so that the
mean value of phase becomes zero
$\left\langle\phi(\textbf{r},t)\right\rangle=0$. Therefore the
order parameter becomes zero too
$\left\langle\Delta(\textbf{r},t)\right\rangle=|\Delta|\left\langle\phi(\textbf{r},t)\right\rangle=0$.
From the above it leads that the asymptotic (\ref{2.19}) at
$T\rightarrow\infty$ is impossible in $2D$ due the phase
fluctuations.

For $D=3$ it is easy to show
\begin{equation}\label{3.8}
\left\langle\Delta^{+}(0)\Delta(\textbf{r})\right\rangle=|\Delta|^{2}
\exp\left[-\frac{k_{B}T}{4\pi^{2}\hbar^{3}c|\Delta|^{2}}\left(q_{c}-\frac{\hbar}{r}\texttt{Si}\left(\frac{rq_{c}}{\hbar}\right)\right)\right].
\end{equation}
Let us consider limit cases:
\begin{equation}\label{3.9}
\begin{array}{ccc}
  r=0 &  &  \left\langle\Delta^{+}(0)\Delta(\textbf{r})\right\rangle=|\Delta|^{2}e^{0}\\
   &  &  \\
  r\rightarrow\infty &  & \left\langle\Delta^{+}(0)\Delta(\textbf{r})\right\rangle=|\Delta|^{2}
  \exp\left[-\frac{k_{B}T}{4\pi^{2}\hbar^{3}c|\Delta|^{2}}\cdot q_{c}\right]>0 \\
\end{array}.
\end{equation}
Thus the long-range order occurs in this case. In Section
\ref{types} in the regime of withholding of the energy gap (type
II solution) we could see the gap (\ref{2.11}) tends to zero
asymptotically $\Delta(q\rightarrow\infty)\rightarrow0$, thus
$q_{c}=\infty$. In this case the integral (\ref{3.4}) over $q$ has
an \emph{ultraviolet} divergence. We can see from
Eqs.(\ref{3.6},\ref{3.9}) that
$\left\langle\Delta^{+}(0)\Delta(\textbf{r})\right\rangle=0$ at
all $r\neq 0$, thus long-rage and quasi long-rage ordering are
absent. In such system despite the fact that the energy gap is
nonzero $|\Delta|\neq 0$ the mean value of the order parameter is
zero $\langle\Delta\rangle=0$, hence superconducting response of
the system is absent. This reflects the fact that superconducting
order parameter is wave function of the condensate of Cooper pairs
$|\Delta|e^{\phi(\textbf{r},t)}$, but due the phase fluctuations
we have $\left\langle\phi(\textbf{r},t)\right\rangle=0$. In the
regime of renormalization (type II solution) the state with the
functions (\ref{2.16}) is normal due the same reasons - the gap
$\upsilon/2$ cannot be destroyed by any finite the condensate's
momentum $q$, that is $q_{c}=\infty$. In contrast, the
superconducting state (\ref{2.18}) has the finite critical
momentum (\ref{2.20}):
$q_{c}\neq\infty\Rightarrow\left\langle\Delta^{+}(0)\Delta(r\rightarrow\infty)\right\rangle\neq0$.
It should be noticed that in the GL theory for $T\rightarrow
T_{c}$ we have $q_{c}\rightarrow 0$ and $|\Delta|\rightarrow 0$ -
Eq.(\ref{2.9}), so that phase fluctuations evolve due suppression
of the gap only (in normal state the phase has not any sense) and
we have the fermionic scenario of the transition. On the contrary
in the regime of renormalization for $T\rightarrow \infty$ we have
$q_{c}\rightarrow \infty$ and $|\Delta|\rightarrow 0$ -
Eq.(\ref{2.20}), so that
$\left\langle\Delta^{+}(0)\Delta(\textbf{r})\right\rangle\rightarrow
0$ i.e. superconducting state is being suppressed by phase
fluctuations (the bosonic scenario).

The ultraviolet limit corresponds to small spatial scales (for
example the coherence length (\ref{2.21}) tends to zero at
$T\rightarrow\infty$). This means that the scale of spatial
variation of the order parameter can be less than interatomic
distance. The similar phenomenon takes place in HeII, where vortex
rings can have minimal radius $\sim5\texttt{A}$ \cite{ray1,car}
with the core radius $\sim 1\texttt{A}$ \cite{ray2} that is less
than average distance between atoms $\sim 3.6\texttt{A}$. As noted
in \cite{tom} such small sizes of the core radius demonstrate that
the vortexes in HeII (and in superconductors) is vortexes in the
field of amplitude of probability $\Psi(\textbf{r})$, but not in
the atomic environment (or in electron gas), and classic formulas
on such scale has effective sense. Moreover in our opinion to cut
off the integrals by some atomic size is not quite correct because
Hamiltonian (\ref{1.2}) is operator in continuous and infinite
medium. It should be noticed that the infrared limit $q\rightarrow
0$ is some abstraction too because it is necessary presence of
infinite medium but in reality we have finite samples.

\section{The assemble of topological defects in the ultraviolet limit $q\rightarrow\infty$}\label{vortex}

As noted above due the infrared divergence in Eq.(\ref{3.4}) the
$1D$ superconductivity is impossible, in $2D$ system BKT
transition occurs and the asymptotic (\ref{2.19}) at
$T\rightarrow\infty$ is impossible. Physically this means
destruction of phase coherence so that
$\left\langle\Delta(\textbf{r})\right\rangle=0$. Mechanism of the
destruction is spontaneous creation of topological defects as the
phase slip ($1D$) and vortexes ($2D$). However we could see that
the ultraviolet divergence in Eq.(\ref{3.4}) takes place in the
regime of withholding of the energy gap due infinity large
critical momentum $q_{c}\rightarrow\infty$. We should investigate
mechanism of the destruction of phase coherence in this case.

Let us calculate the second critical field $H_{c2}$ for the regime
of withholding of the energy gap. To do it a simple method of
calculation of the field $H_{c2}$ can be proposed by the example
of the GL theory. Let a superconductor is in magnetic field
$\textbf{H}\upuparrows Oz$. It is convenient to choose a
calibration $A_{y}=\mu_{0}Hx$. Then GL equation has a form
\cite{tinh}:
\begin{equation}\label{4.1}
    \xi^{2}\left[-\frac{\texttt{d}^{2}}{\texttt{d}x^{2}}+\frac{2\pi i}{\Phi_{0}}\mu_{0}Hx\frac{\texttt{d}}{\texttt{d}y}
    +\left(\frac{2\pi\mu_{0}H}{\Phi_{0}}\right)^{2}x^{2}\right]\varphi-\varphi+|\varphi|^{2}\varphi=0,
\end{equation}
where $\varphi(\textbf{r})=\Delta(\textbf{r})/(a/b)$ is a
dimensionless variable. Here, unlike the standard method, we
retained a term $\varphi^{3}$. When the field strength is of about
$H_{c2}$ the superconductor are pierced by many vortices, so that
the order parameter is strongly nonhomogeneous
$\varphi=\varphi(x,y)$, it varies over distances of the order of
the coherence length $\xi$. We can consider the order parameter is
real $\varphi=\varphi^{+}$ and average it over the system so that
$\langle\varphi(x,y)\rangle=\varphi=\texttt{const}>0$. In
addition, we can suppose $x^{2}=\xi^{2}$. Then Eq.(\ref{4.1})
takes the form:
\begin{equation}\label{4.2}
    \xi^{4}\left(\frac{2\pi\mu_{0}
    H}{\Phi_{0}}\right)^{2}\varphi-\varphi+\varphi^{3}=0,
\end{equation}
where $\Phi_{0}$ is a magnetic flux quantum. Then the order
parameter is
\begin{equation}\label{4.3}
    \varphi=\sqrt{1-\xi^{4}\left(\frac{2\pi\mu_{0}H}{\Phi_{0}}\right)^{2}}.
\end{equation}
We can see $\varphi$ decreases with increasing magnetic field. At
the field
\begin{equation}\label{4.4}
    H_{c2}=\frac{\Phi_{0}}{2\pi\mu_{0}\xi^{2}}
\end{equation}
the second order phase transition takes place $\varphi(H_{c2})=0$.
At $H>H_{c2}$ superconducting phase is absent. Besides the gap
$\varphi(\textbf{r})$ in a vortex can be approximated by the
formula $\varphi(r)=\tanh\left(\nu r/\xi\right)$ \cite{tinh},
where $\nu$ is a constant or order of one. Thus at small $r$ we
have $\varphi\propto r$. However near the vortex line $r<\xi$ the
energy gap is filled with quasiparticles with energy scale
$|\Delta|^{2}/E_{F}\ll |\Delta|$ \cite{genn}. Thus the vortex line
is equivalent to a normal region of radius $\xi$. Let a magnetic
flux is $\Phi=\mu_{0}HS=N\Phi_{0}$, square per one vortex is
$S/N=\Phi_{0}/\mu_{0}H$. The square of one vortex is $\pi\xi^{2}$
Then for $H=H_{c2}$ we have $S/N=2\pi\xi^{2}$. Thus the second
critical field is such field at which the average distance between
the vortices is $\sim\xi$. In other worlds vortexes cannot be
closed to each other by a distance less than $\sim\xi$ because it
leads to destruction of superconducting state.

Using the foregoing method and the functional (\ref{2.10}) we can
write Eq.(\ref{4.2}) for regime of withholding of energy gap:
\begin{equation}\label{4.5}
    \xi^{4}\left(\frac{2\pi\mu_{0}
    H}{\Phi_{0}}\right)^{2}\varphi+\varphi-\frac{\varphi}{|\varphi|}=0.
\end{equation}
Then the gap is
\begin{equation}\label{4.6}
    \varphi=\frac{1}{1+\xi^{4}\left(\frac{2\pi\mu_{0}H}{\Phi_{0}}\right)^{2}}.
\end{equation}
We can see $\varphi$ decreases with increasing magnetic field,
however, unlike GL regime and the regime of renormalization, we
have that
\begin{equation}\label{4.7}
    H_{c2}=\infty.
\end{equation}
This means that any field cannot destroy the energy gap in the
system. In such system the vortex lines $r=0$ is normal only. The
vortexes can be closed to each other arbitrarily close without
destruction of state with energy gap. This fact leads to following
important physical effect.

\begin{figure}[h]
\includegraphics[width=6cm]{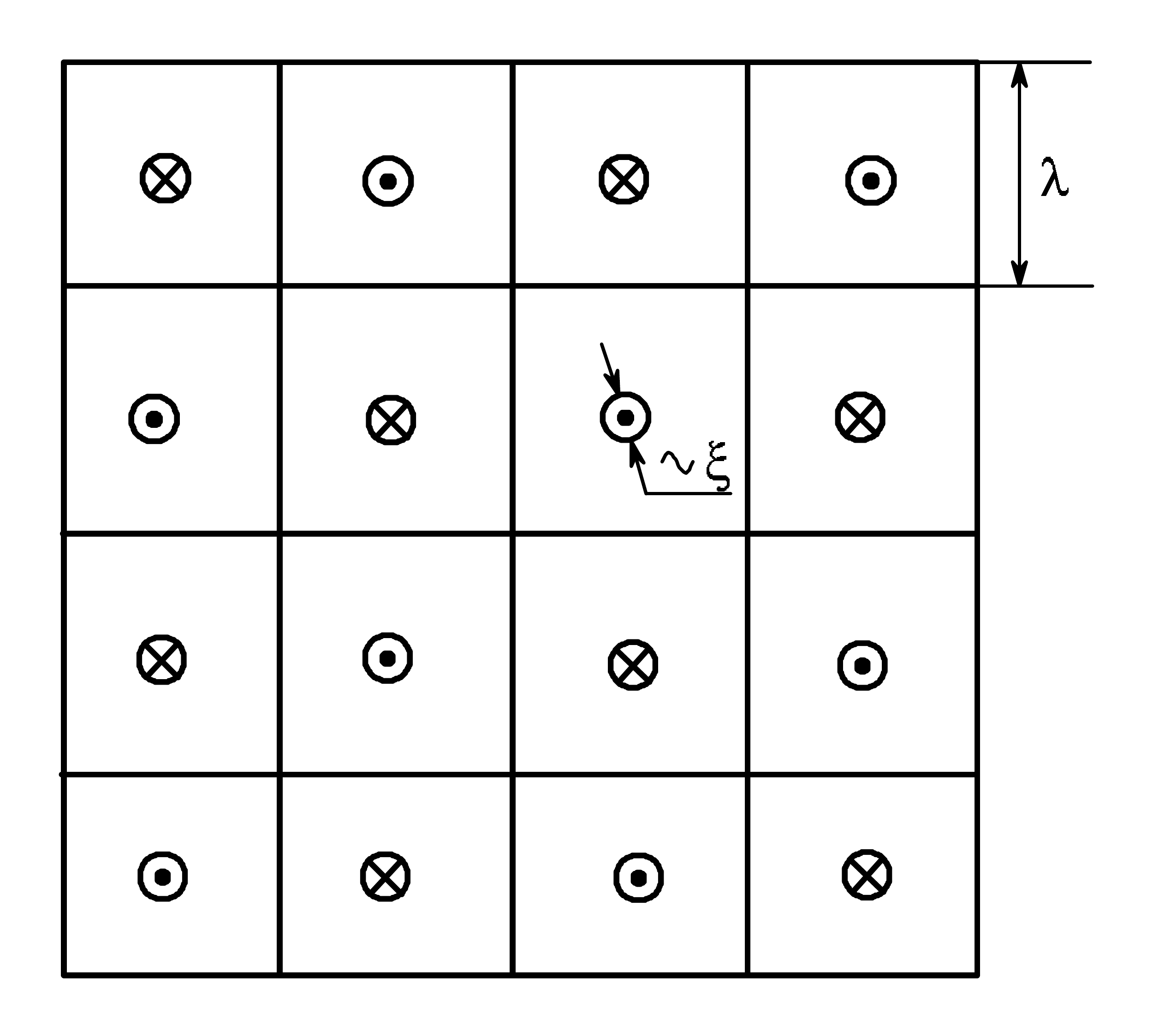}
\caption{Every vortex occupies a cell of the size of magnetic
penetration depth $\lambda$ in order to move independently of
oppositely directed neighbor. The number of ways to place a vortex
in the cell is $\lambda^{2}/\xi^{2}$ since the vortex occupies
square $\sim\xi^{2}$.} \label{Fig2}
\end{figure}

Let us consider creation of a pair of vortexes with opposite
momentums in a $2D$ system. The energy of a vortex is
\begin{equation}\label{4.8}
    \epsilon=d\frac{\Phi_{0}}{8\pi\mu_{0}}B(0)=\left(\frac{\Phi_{0}}{4\pi\lambda}\right)^{2}\frac{d}{\mu_{0}}\ln\frac{\lambda}{\xi},
\end{equation}
where $B(0)$ is magnetic induction in center of the vortex,
$\lambda$ is a magnetic penetration depth, $d$ is layer's
thickness. Interaction between the vortexes is
$U(r)=-\frac{\Phi_{0}}{8\pi^{2}\lambda^{2}}\ln\frac{\lambda}{r}$,
thus at distances $r>\lambda$ the vortexes do not interact almost
and they behave as free from each other. However in other parts of
the system the same process take place. Thus the system can be
divided into cells of size $\sim\lambda$ of each - Fig.\ref{Fig2}.
The number of ways to place a vortex in the cell is
$\lambda^{2}/\xi^{2}$ since the vortex occupies square
$\sim\xi^{2}$. Corresponding entropy is
$k_{B}\ln(\lambda^{2}/\xi^{2})$.  Then free energy per a vortex is
\begin{equation}\label{4.10}
    F=\epsilon-TS=\frac{d}{\mu_{0}}\left(\frac{\Phi_{0}}{4\pi\lambda}\right)^{2}\ln\frac{\lambda}{\xi}-k_{B}T\ln\frac{\lambda^{2}}{\xi^{2}}
    =\left[\frac{d}{\mu_{0}}\left(\frac{\Phi_{0}}{4\pi\lambda}\right)^{2}-2k_{B}T\right]\ln\frac{\lambda}{\xi}.
\end{equation}
Free vortexes can appear when $F=0$, that is at temperature
\begin{equation}\label{4.11}
    k_{B}T_{\texttt{BKT}}=\frac{d\Phi_{0}^{2}}{32\pi^{2}\mu_{0}\lambda^{2}}.
\end{equation}
The phase of order parameter $\phi(\textbf{r},t)$ in each point of
superconductor depends strongly (in each time moment) on position
all vortexes because the phase changes to $2\pi$ at going around
each vortex. Moving of the free vortexes leads to strong
fluctuations of the phase $\phi(\textbf{r},t)$ so that the mean
value of phase becomes zero
$\left\langle\phi(\textbf{r},t)\right\rangle=0$, hence the order
parameter becomes zero too
$\left\langle\Delta(\textbf{r},t)\right\rangle=|\Delta|\left\langle\phi(\textbf{r},t)\right\rangle=0$.

As we have seen above for regime of withholding of energy gap the
system is characterized by the finite coherence length $\xi$
(\ref{2.11a}), but in a vortex the line $r=0$ is normal only,
unlike the GL theory where normal core has radius $\sim\xi$. Thus
the energy of a vortex remains finite value because it is
determined by magnetic induction $B(0)$ in center of the vortex,
however the number of ways to place a vortex in the cell is
$\lambda^{2}/0$ since the vortex occupies zero square. Then the
free energy per a vortex is
\begin{equation}\label{4.10}
    F=\epsilon-TS=\frac{d}{\mu_{0}}\left(\frac{\Phi_{0}}{4\pi\lambda}\right)^{2}\ln\frac{\lambda}{\xi}-k_{B}T\ln\frac{\lambda^{2}}{0^{2}}
    =-\infty.
\end{equation}
Thus we can make the following conclusion: \emph{creation of a
topological defect} (vortex in $2D$ system or vortex ring, vortex
tangle in $3D$ or other forms in quantum turbulence) \emph{in the
regime of withholding of energy gap is accompanied by the
unlimited increase of entropy but it's energy remains finite}.
This means that the superconducting state is unstable and
destroyed by phase fluctuations so that the order parameter is
zero $\langle\Delta(\textbf{r},t)\rangle=|\Delta|\langle
e^{\phi}(\textbf{r},t)\rangle=0$ but energy gap can be nonzero
$|\Delta(\textbf{r},t)|\neq 0$ like in the pseudogap state.

\section{Summary}\label{summ}

In this article we investigate the stability of superconducting
state in the BCS model with the external pair potential. This
model has two types of solutions: regime of withholding of the
energy gap and regime of renormalization. The first type has
property: if even the coupling constant is $\lambda=0$ then the
energy gap is nonzero $|\Delta|=\upsilon/2$ and it does not depend
on temperature, moreover the gap tends to zero asymptotically with
momentum of the condensate $q$ unlike GL theory where the critical
momentum $q_{c}$ exists such that $\Delta(q>q_{c})=0$. Thus EPP
imposes superconductivity to the system and holds it hardly at all
temperatures and on all spatial scales. The second type solution
has property: if the coupling constant is $\lambda=0$ then
$\Delta=0$. This means, that the potential $\upsilon$ renormalizes
the order parameter $\Delta$ only but it is not the cause of
superconductivity. At large temperatures $T\gg T_{c}(\upsilon=0)$
the energy gap tends to zero asymptotically as $1/T$. Thus,
formally, the critical temperature is equal to infinity. However
the critical momentum $q_{c}$ exists like in ordinary GL theory.
In this regime if we suppose the ordering $\Delta\sim\left\langle
aa\right\rangle, \Delta^{+}\sim\left\langle
a^{+}a^{+}\right\rangle$ is absent then the energy gap
$\upsilon/2$ presents in the quasiparticles' specter. We suppose
that the state with $\Delta=0$ can be interpreted as normal state
with a pseudogap, i.e. current carriers are the pairs but the
pairs are not correlated. The pseudogap state cannot be destroyed
by the finite condensate's momentum $q$ as in previous regime.

We have shown that if EPP is present $\upsilon>0$ but
electron-electron interaction is absent $g=0$ that
superconductivity is impossible:
$\left\langle\Delta^{+}(0)\Delta(\textbf{r})\right\rangle_{r\rightarrow\infty}=0$
although $|\Delta(\textbf{r})|\neq0$. This is result of phase
fluctuations such that
$\left\langle\Delta(\textbf{r},t)\right\rangle=|\Delta(\textbf{r})|\left\langle
e^{i\phi(\textbf{r},t)}\right\rangle=0$ i.e. the superconducting
ordering is destroyed by these fluctuations. Thus the regime  of
withholding of the energy gap and the normal state of the regime
of renormalization is the pseudogap state: there are stable but
noncoherent Cooper pairs in the system. This means that \emph{we
cannot impose the superconducting state on the noninteracting
system by the external pair potential}, only electron-electron
interaction is cause of superconductivity. We can impose the
superconducting state on the noninteracting system in a case only
when the operator $\widehat{H}_{\upsilon}$ has $U(1)\times U(1)$
symmetrical form $\sum_{\textbf{k}}\left[\upsilon
a_{\textbf{k}\uparrow}^{+}a_{-\textbf{k}\downarrow}^{+}+\upsilon^{+}
a_{-\textbf{k}\downarrow}a_{\textbf{k}\uparrow}\right]$, where
$\upsilon$ is the order parameter of another superconductor, for
example, the boundary (proximity) effect, when a superconductor is
placed in contact with a normal metal \cite{genn} or interband
mixing of two order parameters belonging to different bands in a
multi-band superconductor \cite{asker,grig1,litak} occurs. In our
model of superconductivity (second type solution) the EPP can
renormalize the order parameter and normal state only: the
asymptotic $\Delta\propto\frac{\upsilon}{T}$ and energy gap in
normal state are effects of this renormalization. Physically these
results is consequence of the fact that the order parameter is a
wave function $|\Delta(\textbf{r})|e^{i\phi(\textbf{r})}$ of the
Cooper pairs but not the energy gap $|\Delta(\textbf{r})|$. This
means that in the BCS model with EPP the bosonic scenario of
suppression of superconducting state occurs - the normal state is
a state with noncorrelated pairs, unlike BCS theory where
fermionic scenario (break of the pairs) takes place.

As well known the impossibility of $1D$ superconductivity and BKT
transition in $2D$ systems is mathematical consequence of the
infrared divergence in the integral of the correlator (\ref{3.4}).
Physically this means destruction of the superconducting ordering
by phase fluctuations. Thus the asymptotic (\ref{1.5}) cannot be
realized in the $2D$ system. For the regime of withholding of the
energy gap and normal state of the regime of renormalization the
ultraviolet divergence occurs as a result of stability of the
energy gap with respect to the condensate's momentum, i.e.
$q_{c}=\infty$ unlike GL theory and the regime of renormalization
where the critical momentum is finite. The ultraviolet limit
corresponds to small spatial scales, to cut off the integrals by
some atomic size is not correct because Hamiltonian (\ref{1.2}) is
operator in continuous and infinite medium, and, for example, the
vortexes is vortexes in the field of amplitude of probability
$\Psi(\textbf{r})$, but not in the electron gas, and classic
formulas on such scale has effective sense. We have found
mechanism of the destruction of phase coherence in this case:
creation of a topological defect (vortex in $2D$ system or vortex
ring, vortex tangle in $3D$ and other forms of quantum turbulence)
in the regime of withholding of energy gap is accompanied by the
unlimited increase of entropy but it's energy remains finite. The
infinity entropy of the vortex is a result of the fact that in the
vortex the line $r=0$ is normal only unlike the GL theory where
normal core has radius $\sim\xi$. Thus the energy of a vortex
remains finite value because it is determined by magnetic
induction $B(0)$ in center of the vortex, however the number of
ways to place a vortex is infinity since its normal core occupies
zero square.

\end{document}